\documentclass[preprintnumbers,superscriptaddress,amsmath,amssymb,nofootinbib,twocolumn,showpacs]{revtex4-1}

\usepackage{dcolumn}
\usepackage{bm}
\usepackage{graphicx}
\usepackage{subcaption}

\usepackage{color}
\usepackage{natbib}
\usepackage{twoopt}
\newcommand{\beq}{\begin{equation}}
\newcommand{\eeq}{\end{equation}}
\newcommand{\ben}{\begin{eqnarray}}
\newcommand{\een}{\end{eqnarray}}
\newcommand{\bi}{\begin{itemize}}
\newcommand{\ei}{\end{itemize}}

\newcommand{\eg}{\mbox{\it e.g.}}

\newcommand{\citesec}[1]{Sect.~\ref{#1}}

\newcommand{\citefig}[1]{Fig.~\ref{#1}}

\newcommand{\fsol}{{\ifmmode f_{\odot} \else $f_{\odot}$\fi}}

\newcommand{\sqgev}{{\ifmmode {\rm GeV}^2 \else ${\rm GeV}^2$\fi}}

\newcommand{\sr}{{\ifmmode {\rm sr} \else ${\rm sr}$\fi}}
\newcommand{\invsr}{{\ifmmode {\rm sr}^{-1} \else ${\rm sr}^{-1}$\fi}}

\newcommand{\scnd}{{\ifmmode {\rm s} \else ${\rm s}$\fi}}
\newcommand{\invscnd}{{\ifmmode {\rm s}^{-1} \else ${\rm s}^{-1}$\fi}}

\newcommand{\kpc}{{\ifmmode {\rm kpc} \else ${\rm kpc}$\fi}}
\newcommand{\invkpc}{{\ifmmode {\rm kpc}^{-1} \else ${\rm kpc}^{-1}$\fi}}

\newcommand{\sqkpc}{{\ifmmode {\rm kpc}^{2} \else ${\rm kpc}^{2}$\fi}}
\newcommand{\invsqkpc}{{\ifmmode {\rm kpc}^{-2} \else ${\rm kpc}^{-2}$\fi}}

\newcommand{\cm}{{\ifmmode {\rm cm} \else ${\rm cm}$\fi}}
\newcommand{\invcm}{{\ifmmode {\rm cm}^{-1} \else ${\rm cm}^{-1}$\fi}}

\newcommand{\sqcm}{{\ifmmode {\rm cm}^2 \else ${\rm cm}^2$\fi}}
\newcommand{\invsqcm}{{\ifmmode {\rm cm}^{-2} \else ${\rm cm}^{-2}$\fi}}

\newcommand{\meter}{{\ifmmode {\rm m} \else ${\rm m}$\fi}}
\newcommand{\invmeter}{{\ifmmode {\rm m}^{-1} \else ${\rm m}^{-1}$\fi}}

\newcommand{\sqmeter}{{\ifmmode {\rm m}^2 \else ${\rm m}^2$\fi}}
\newcommand{\invsqmeter}{{\ifmmode {\rm m}^{-2} \else ${\rm m}^{-2}$\fi}}

\newcommand{\lcdm}{{\ifmmode \Lambda{\rm CDM} \else $\Lambda{\rm CDM}$\fi}}

\newcommand{\Rvirh}{{\ifmmode R_{\rm vir}^{\rm h} \else 
    $R_{\rm vir}^{\rm h}$\fi}}

\newcommand{\Ncl}{{\ifmmode N_{\rm cl} \else $N_{\rm cl}$\fi}}
\newcommand{\ncl}{{\ifmmode n_{\rm cl} \else $n_{\rm cl}$\fi}}
\newcommand{\phicl}{{\ifmmode \phi_{\rm cl} \else $\phi_{\rm cl}$\fi}}
\newcommand{\phicltot}{{\ifmmode \phi_{\rm cl}^{\rm tot} \else 
    $\phi_{\rm cl}^{\rm tot}$\fi}}
\newcommand{\Beff}{{\ifmmode B_{\rm eff} \else $B_{\rm eff}$\fi}}

\newcommand{\aap}{Astron.~Astroph.}

\newcommand{\apjl}{Astrophys.~J.~Lett.}
\newcommand{\apss}{Astrophysics and Space Science}

\newcommand{\jgr}{J.~Geophys.~Res.}

\newcommand{\jcap}{JCAP}
\newcommand{\mnras}{MNRAS}
\newcommand{\pasj}{Pub.~Astron.~Soc.~Jap.}

\newcommand{\ssr}{Space Science Reviews}

\usepackage[pdftex,
            breaklinks=true,%
            colorlinks=true,%
            linkcolor=red,
            urlcolor=blue,
            citecolor=blue,
            pdfauthor={Stref, Lavalle},%
            pdftitle={Template for article in XXX}%
           ]{hyperref}

\graphicspath{{./}{Figs/}}

\begin{document}

\title{Novel Cosmic-Ray Electron and Positron Constraints on MeV Dark Matter Particles}
\author{Mathieu Boudaud}
\email{boudaud@lpthe.jussieu.fr}
\affiliation{LAPTh, Universit\'e Savoie Mont Blanc \& CNRS, 9 Chemin de Bellevue, B.P.110,
  F-74941 Annecy-le-Vieux -- France}
\affiliation{Laboratoire de Physique Th\'eorique et Hautes \'Energies (LPTHE),
  UMR 7589 CNRS \& UPMC, 4 Place Jussieu, F-75252 Paris -- France}

\author{Julien Lavalle}
\email{lavalle@in2p3.fr}
\affiliation{Laboratoire Univers \& Particules de Montpellier (LUPM),
  CNRS \& Universit\'e de Montpellier (UMR-5299),
  Place Eug\`ene Bataillon,
  F-34095 Montpellier Cedex 05 -- France}

\author{Pierre Salati}
\email{pierre.salati@lapth.cnrs.fr}
\affiliation{LAPTh, Universit\'e Savoie Mont Blanc \& CNRS, 9 Chemin de Bellevue, B.P.110,
  F-74941 Annecy-le-Vieux -- France}

\begin{abstract}
  MeV dark matter (DM) particles annihilating or decaying to electron-positron pairs cannot, in
  principle, be observed via local cosmic-ray (CR) measurements because of the shielding solar
  magnetic field. In this letter, we take advantage of spacecraft Voyager 1's capacity for
  detecting interstellar CRs since it crossed the heliopause in 2012. This opens up a new avenue
  to probe DM in the sub-GeV energy/mass range that we exploit here for the first time. From a
  complete description of the transport of electrons and positrons at low energy, we derive
  predictions for both the secondary astrophysical background and the pair production mechanisms
  relevant to DM annihilation or decay down to the MeV mass range. Interestingly, we show that
  reacceleration may push positrons up to energies larger than the DM particle mass. We combine
  the constraints from the Voyager and AMS-02 data to get novel limits covering a very extended
  DM particle mass range, from MeV to TeV. In the MeV mass range, our limits reach annihilation
  cross sections of order $\langle \sigma v\rangle \sim 10^{-28}{\rm cm^3/s}$. An
  interesting aspect is that these limits barely depend on the details of cosmic-ray propagation
  in the weak reacceleration case, a configuration which seems to be favored by the most
  recent boron-to-carbon ($B/C$) data.
  Though extracted from a completely different and new probe, these bounds have a strength
  similar to those obtained with the cosmic microwave background --- they are even more
  stringent for $p$-wave annihilation.
\end{abstract}

\maketitle
\preprint{LUPM:16-022}

Thermally produced sub-GeV dark matter (DM) particles have triggered interest since the
nonbaryonic particle DM proposal itself, including the weakly interacting massive particle (WIMP)
paradigm (\eg~\cite{Bond1982}). Allowed scenarios involve DM particle masses $m_\chi$ larger
than a few keV (warm DM -- WDM), usually bounded by structure formation
\cite{ColombiEtAl1996,Viel2005,Viel2013,Bose2016}. In the MeV mass range, thermal DM candidates
are already cold enough not to differ from cold DM (CDM) structure formation on scales of dwarf
galaxies. However, if they have remained coupled to radiation or neutrinos sufficiently
long, an oscillatory damping pattern in the structure power spectrum could be observed on
small scales that differs from the standard free-streaming cutoff of WDM
\cite{BoehmEtAl2004,BoehmEtAl2014}, and alleviate the too-big-to-fail problem affecting the CDM
paradigm on small scales \cite{Boylan-Kolchin2011}. Besides, self-interacting DM scenarios could
be achieved from the thermal freeze out of particles in the MeV mass range \cite{ChuEtAl2016}, as
well as strongly interacting DM \cite{HochbergEtAl2014}, which may also cure small scale issues in
structure formation. Overall, many efforts are now devoted to probe this mass range
through direct and indirect searches (see {\em e.g.} \cite{AlexanderEtAl2016}).

Astrophysical observations already constrain MeV DM candidates. For instance, gamma-ray
measurements generically constrain MeV candidates, depending on assumptions on the shape of the
inner Galactic halo profile \cite{Beacom2005,Essig2013a}. Heating of the plasma at the CMB
decoupling time is also constrained by current observations and actually allows us to get stringent
bounds on MeV annihilating or decaying DM
\cite{Chen2004,Slatyer2016,LiuEtAl2016}, down to an $s$-wave annihilation cross section of
$\langle \sigma v \rangle\lesssim 10^{-29}\,{\rm cm^3/s}$ for the former case (assuming
annihilation into electron-positron [$e^{\pm}$] pairs).

Less prone to uncertainties in the halo shape \cite{Lavalle2012a}, cosmic-ray (CR) $e^{\pm}$s
could provide independent probes of annihilation or decay of MeV DM.
Nevertheless, interstellar sub-GeV $e^{\pm}$s are shielded by the solar magnetic
field (the so-called solar modulation effect) \cite{Gleeson1968a,Potgieter2013} such that they
cannot reach detectors orbiting the Earth. In this Letter, we bypass this limitation by
exploiting, for the first time in this context (see \eg\ \cite{CummingsEtAl2016} for
  more conventional astrophysical aspects), the $e^{\pm}$ data of the Voyager 1 spacecraft
\cite{KrimigisEtAl1977,StoneEtAl2013}. Indeed, Voyager 1 has crossed the heliopause during the
summer 2012, and, since then, has traveled through interstellar space. Since it is
equipped with particle detectors, with one dedicated to $e^{\pm}$ measurements (no discrimination
between electrons and positrons), this opens up a new avenue for DM searches in the sub-GeV mass
range. Here, we will use the $e^{\pm}$ Voyager 1 data from the end of 2012, extracted after the
calibration of response functions from simulations of the detector (most conservative dataset) and
released in Ref.~\cite{StoneEtAl2013} -- {\em i.e.} 4 data points in the $\sim 10-50$ MeV energy
range with excellent statistics. This data set will be complemented at higher energy by the AMS-02
positron data \cite{Aguilar2014}, imported from the database proposed in \cite{MaurinEtAl2014}.

The transport of CR $e^{\pm}$s in the Milky Way (MW) can be described by a general diffusion
equation \cite{Ginzburg1964a,Bulanov1974,Moskalenko1998,Delahaye2009a,Delahaye2010,Boudaud2015,BoudaudEtAl2016} that includes spatial diffusion, convection, reacceleration, and energy losses. We use
the semianalytic method proposed in Ref.~\cite{Maurin2001} to solve this equation. Solutions
assume a plain diffusion over a cylindrical magnetic halo of radius $R$ and half height $L$, with
boundary conditions such that the CR density vanishes at the halo borders. Some processes are
dominant in the thin Galactic disk, others extend to the whole magnetic halo. The first category
includes diffusive reacceleration (featured by a pseudo-Alfv\'en velocity $V_a$), and energy losses
due to electromagnetic interactions with the interstellar gas [$b_{\rm gas}(E)$]. The second one
includes spatial diffusion (with a scalar coefficient $K({\cal R})=\beta \,
K_0({\cal R}/1\,{\rm GV})^\delta$, with ${\cal R}\equiv p/|q|$ the rigidity), convection (with
velocity $\vec{V}_c={\rm sgn}(z)\,V_c\,\vec{e}_z$ of constant modulus), and higher-energy losses
from inverse Compton and synchrotron emissions ($b(E)$). The technical difficulty in applying
this method to $e^{\pm}$s comes from the fact that derivatives in the momentum space are not confined
to the disk as is the case for nuclei, but high-energy losses, which dominate above $\sim 10$ GeV,
are efficient all over the magnetic halo. This was addressed in an approximate way in
Ref.~\cite{Delahaye2009a}, but recently solved in a systematic and
elegant way in Ref.~\cite{BoudaudEtAl2016}. We therefore refer the reader to
Ref.~\cite{BoudaudEtAl2016} for a thorough presentation of the propagation model we will further use
in this Letter. Note that complementary full numerical approaches exist
\cite{Strong1998,Evoli2008,Kissmann2014}, which are qualitatively similar to ours.

For the propagation parameters, we consider large-$L$ propagation models because
low-energy positron data (0.1-2 GeV) severely constrain values of $L\lesssim 8$ kpc
\cite{Lavalle2014,BoudaudEtAl2016}, as do the latest $B/C$ data \cite{Kappl2015a}. More precisely,
we use the Max model proposed in
\cite{Maurin2001,Donato2004} (model $A$ henceforth: $L=15$ kpc,
  $K_0=0.0765\,{\rm kpc^2/Myr}$, $\delta=0.46$, $V_a=117.6\,{\rm km/s}$, $V_c=5\,{\rm km/s}$),
which lies at the border of the current positron bounds \cite{BoudaudEtAl2016}, together with the
$B/C$ best-fit model of Ref.~\cite{Kappl2015a} (model $B$: $L=13.7$ kpc,
  $K_0=0.0967\,{\rm kpc^2/Myr}$, $\delta=0.408$, $V_a=31.9\,{\rm km/s}$, $V_c=0.2\,{\rm km/s}$).
These models mostly differ in reacceleration, which is strong for
model $A$ (fitted on old B/C data), and weak for model $B$ (most recent B/C data)\footnote{The
  relevant (inverse) time scale is given by
  $V_a^2/K_0 \sim 0.2 \,{\rm Myr^{-1}}$ for model $A$ ($0.01\,{\rm Myr^{-1}}$ for model B).}.
A full exploration of the parameter space goes beyond the scope of this Letter, but models
$A$ and $B$ characterize the state-of-the-art description of Galactic CR propagation within a
standard set of assumptions (isotropic and scalar spatial diffusion).
%
%
\begin{figure}[!t]
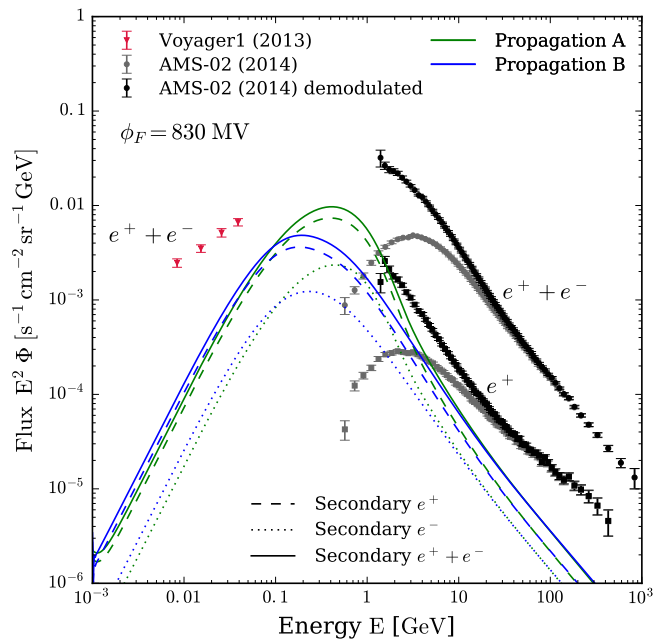

\centering
\includegraphics[width = 0.495\textwidth]{{{electron_positron_flux_II}}}
\caption{\small CR $e^{\pm}$ data from Voyager (red triangles) and AMS-02 (plain circles), and
  $e^+$ AMS-02 data (plain squares). The latter are demodulated with a Fisk potential of
  $\phi=830$ MV. The curves show the interstellar secondary background predicted for propagation
  models $A$ and $B$.}
\label{fig:data_and_secs}
\end{figure}
We first compute the secondary $e^{\pm}$ fluxes, {\em i.e.} $e^{\pm}$s generated from
inelastic interactions between CR nuclei and the interstellar gas. Though conventional sources
of primary CRs ({\em e.g.} pulsar winds, supernova remnants) contribute to the total $e^{\pm}$
fluxes \cite{Shen1970,Aharonian1995,Delahaye2010,Boudaud2015}, we only consider the secondary
background because large theoretical uncertainties affect this primary component
\cite{Delahaye2010,Boudaud2015}, placing our coming constraints on the conservative
side. Our predictions for the interstellar flux are shown in \citefig{fig:data_and_secs} against the
Voyager and AMS-02 data. For the latter, we demodulated the data (the Voyager
data are modulation free). We proceed by using the force-field approximation
\cite{Gleeson1968a,Fisk1971} with a Fisk potential $\phi$ in the range
$[724\,{\rm MV},830\,{\rm MV}]$ for the AMS-02 data-taking period
(see~\cite{GhelfiEtAl2016}).
From \citefig{fig:data_and_secs}, we see that the secondary $e^{\pm}$s
contribute significantly to the data only in the AMS-02 energy range and are negligible in the
Voyager range. This has important consequences: not only are the Voyager data free
of solar modulation, but they are also insensitive to the presence of secondaries. Besides, we can
already notice from \citefig{fig:data_and_secs} the impact of reacceleration: the secondary
$e^{\pm}$ peak observed in $E^2\Phi_{e^\pm}$ is shifted to higher energy in the
strong-reacceleration case (model $A$), which will make the AMS-02 data more constraining than
in the weak-reacceleration case (model $B$). Because confined to the disk here,
  reacceleration in both models is consistent with the power constraints found in \cite{Thornbury2014,DruryEtAl2017}.
Finally, it is worth pointing out that, surprisingly enough, the lowest AMS-02
energy point for the $e^+$ flux lies significantly lower than its neighbors, which may lead to
very strong bound on DM annihilation or decay. To remain conservative, we remove it from our
analysis.
%

We now compute the DM annihilation contributions to these observables. We consider several channels
and assume unity branching ratios: $e^{\pm}$, $\mu^{\pm}$,
$\tau^{\pm}$, $b\bar b$, $W^\pm$. We generate injection spectra with the \texttt{MicrOMEGAs} code
\cite{BelangerEtAl2015}, which includes final-state radiation (FSR) processes. For the DM halo
profile, we assume two different spherical cases: a Navarro-Frenk-White halo \cite{Navarro1997}
scaling like $1/r$ in the center (NFW halo henceforth), and a cored halo profile with constant
central DM density (cored halo). We use the kinematically constrained halo parameters
from Ref.~\cite{McMillan2017}, such
that our halos are dynamically self-consistent. In both halos, the DM density at the solar
position $r_\odot\simeq 8.2$ kpc is $\rho_\odot\simeq 0.4\,{\rm GeV/cm^3} $.

\begin{figure}[!t]
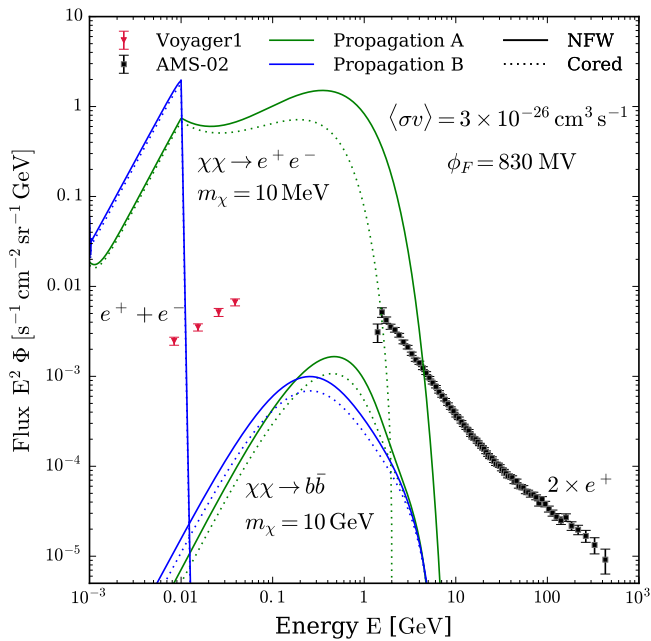

\centering
\includegraphics[width = 0.495\textwidth]{{{electron_positron_flux_I}}}
\caption{\small Predictions ($e^\pm$) for 2 template cases: a 10 MeV WIMP annihilating
  into $e^+e^-$ (+ FSR), and a 10 GeV WIMP annihilating into $b\bar b$. The data are the same as
  in \citefig{fig:data_and_secs}, but the AMS-02 $e^+$ data is multiplied by a factor
  of 2 to compare with the $e^\pm$ primaries. Propagation models $A$ and $B$,
  and the NFW and cored DM halo models were used.}
\label{fig:prims}
\end{figure}

Template predictions for the DM-induced $e^\pm$ fluxes are shown in
\citefig{fig:prims}, considering WIMPs of 10 MeV (10 GeV) annihilating into $e^+ e^-$
($b\bar b$). In both cases, $e^+$s and $e^-$s share the same injection spectrum and the same
propagation history, such that $e^\pm$ predictions can be compared to the $e^+$
data by multiplying the latter by two. We reported our results for propagation models $A$ and $B$,
and for the NFW and cored halos. In the weak-reacceleration case (model $A$), the $e^\pm$ flux is
suppressed beyond the maximal injected energy set by $m_\chi$, while in the strong-reacceleration
case (model $B$), low-energy $e^\pm$s are reaccelerated beyond $m_\chi$. This important feature of
the strong-reacceleration regime has, to our knowledge, never been noticed before: DM-induced
$e^\pm$s could then be observed beyond $m_\chi$, which makes the GeV data also relevant to
constrain sub-GeV DM.

Reacceleration also rules the impact of the DM halo shape. Without reacceleration, sub-GeV
CR propagation is mostly governed by energy losses, such that $e^\pm$s injected at sub-GeV
energies and coming from regions close to the Galactic center (GC) have been drifted to the
low-energy part of the spectrum. This is illustrated in \citefig{fig:prims} for the $e^+e^-$
channel, for which even if an NFW halo induces a larger annihilation rate at the GC, the net
increase in the $E^2\Phi_{e^\pm}$ curve for model $B$ is lost at low energies, while the peak at
the WIMP mass only reflects the very local annihilation rate. For the $b\bar b$
channel, GeV $e^\pm$s injected at GeV energies in the GC are locally observed at sub-GeV energies,
hence a larger flux for a cuspy halo. On the other hand, efficient reacceleration (model $A$)
makes these $e^\pm$s continuously reheated as they cross the disk on their way to us, compensating
for energy losses, such that the difference between an NFW and a cored halo is now more
pronounced beyond $m_\chi$ (though still much less than the differences induced in gamma-ray
predictions \cite{Lavalle2012a}).
This non-trivial effect of reacceleration strengthens the complementarity of the low-energy
Voyager data with the higher-energy AMS-02 data, the former (latter) providing significant
constraints on predictions based upon weak-(strong-)reacceleration models.

The AMS-02 data
are particularly constraining in the strong-reacceleration case, as secondary $e^+$s provide
a large contribution above 100 MeV ($\sim$ the charged pion mass), while more sensitive
to uncertainties in the solar modulation or $V_a$. In contrast, flux predictions in the sub-GeV
range and in the weak-reacceleration regime are almost not sensitive to uncertainties in the
other propagation parameters. This is because ionization energy losses are then the dominant
process in the sub-GeV energy range (see \citesec{app:prop}). The corresponding rate
$b_{\rm ion}$ scales like the local gas density, for which the uncertainties are small
\cite{Ferriere2001,NakanishiEtAl2003}. In this configuration, the peak
observed in $E^2\phi$ at the WIMP mass in the $e^+ e^-$ channel, whose amplitude fixes the
Voyager bound on $\langle \sigma v\rangle$, is predicted to an excellent precision from the
asymptotic approximation $\phi_{e^\pm}(E\to m_\chi)\to
(v/4\pi)(\rho_\odot/m_\chi)^2 \langle \sigma v \rangle /b_{\rm ion}(E\to m_\chi)$.
%
%
\begin{figure*}[!t]
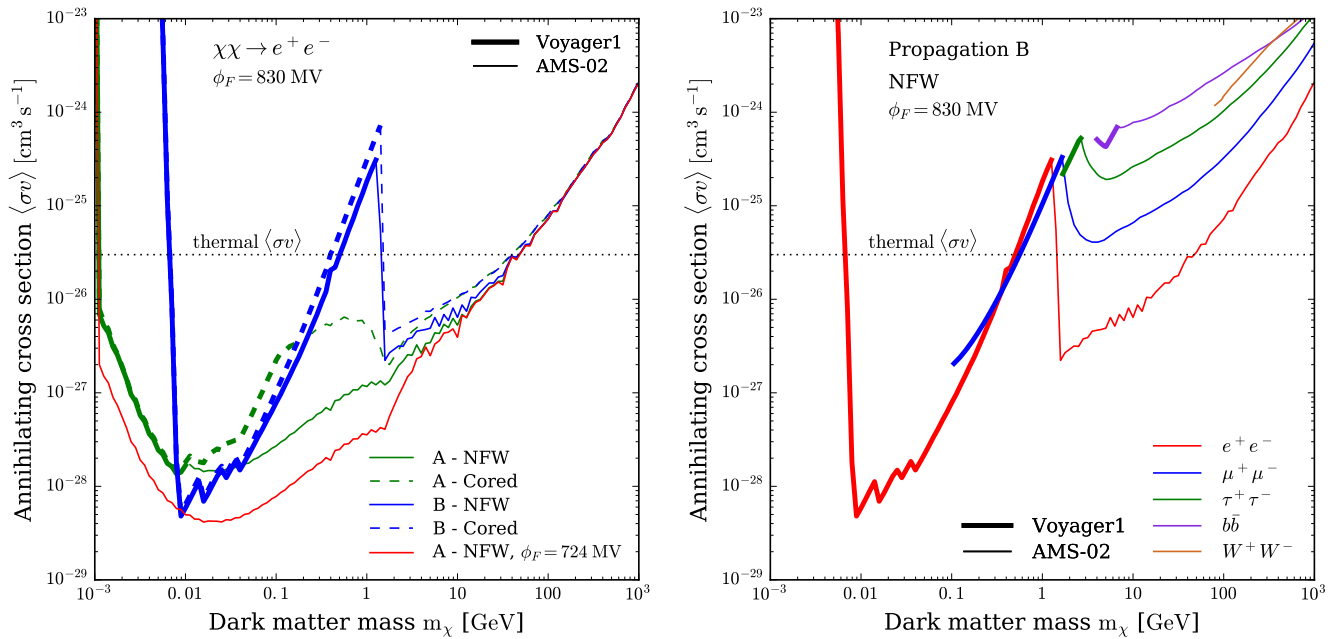

\centering
\includegraphics[width = 0.495\textwidth]{{{sigmav_bounds_uncertainties}}}
\includegraphics[width = 0.495\textwidth]{{{sigmav_bounds_results}}}
\caption{\small Limits on $\langle \sigma v\rangle$ as a function of $m_\chi$. Left: limits
  assuming annihilation into $e^+e^-$ for propagation models $A$ and
  $B$, and for the NFW and cored DM halos. A conservative solar modulation is set with
  $\phi=830$ MV. The result for $\phi=724$ MV is shown for the $A$-NFW configuration.
  Right: limits for different annihilation final states, assuming configuration $B$-NFW-830 MV.}
\label{fig:limits}
\end{figure*}

We now combine the Voyager and AMS-02 data discussed above to derive limits on
$\langle \sigma v\rangle$. We assume Majorana DM particles -- a factor of 2 must be applied to our
limits for Dirac fermions. We also assume that $\langle \sigma v\rangle$ is position independent
(valid for an $s$-wave, approximate for a $p$-wave). We derive limits by adding our flux predictions
for the primary and secondary components, and then demanding the total flux to lie below 2$\sigma$
from each data point. These limits are displayed in \citefig{fig:limits}. In the left panel, we
specialize to the $e^+e^-$ channel to illustrate differences due to propagation, solar modulation,
and the DM halo shape. As already emphasized, the main variation is driven by reacceleration:
weak-reacceleration models ($\sim$ model $B$) are severely constrained by the Voyager data below
$\sim 100$ MeV, with the nice bonus of not suffering from solar modulation. On the other hand,
this regime makes it possible to ``hide'' a positron $E^2\Phi_{e^+}$ peak in the blind spot between
the Voyager and AMS-02 datasets, such that the 0.1-1 GeV mass range becomes unconstrained. In
contrast, strong-reacceleration models ($\sim$ model $A$) forbid any blind spot,
simply because sub-GeV $e^\pm$s are shifted up to GeV energies, in which case AMS-02
constraints are turned on -- curves get then smoother over the full MeV-TeV energy range,
with a transition below $\sim 10$ MeV where Voyager takes over. Limits inferred from
strong-reacceleration models are also more sensitive to solar modulation, as illustrated by
decreasing $\phi=830$ MV to 724 MV: this justifies our conservative choice of 830 MV.

In the right panel of \citefig{fig:limits}, we generalize our limits for several
annihilation channels conservatively assuming propagation model $B$, $\phi=830$ MV for the solar
modulation of the AMS-02 data, and the NFW halo (closer to the best fit of \cite{McMillan2017}
than the cored halo). These are our main results, which demonstrate for the first time that
CR $e^\pm$s constrain annihilating DM down to the MeV mass range. We emphasize that for the
$e^+e^-$ channel our bound reaches $\langle \sigma v\rangle \sim 10^{-28}{\rm cm^3/s}$ in the
10-100 MeV mass range. We also notice the blind spot just below 1 GeV, but we stress that more
reacceleration would fill in this spot again -- future studies on propagation parameters will
be crucial to settle this. At higher energy, we exclude thermal cross
sections ($\sim 3\times 10^{-26}{\rm cm^3/s}$) for masses up to $\sim 50$ GeV. This is less
stringent than bounds obtained in Ref.~\cite{Bergstroem2013}, where the authors have assumed
an additional primary component from pulsars that saturates the data and forbids DM-induced
contributions. Because of the large uncertainties affecting
this primary component \cite{Delahaye2010,Boudaud2015}, we have instead decided to discard it.

We now compare our results with those obtained from CMB analyses. In Ref.~\cite{LiuEtAl2016},
limits on $s$-wave annihilation obtained for the $e^+e^-$ channel go from
$\langle \sigma v\rangle \lesssim 3\times 10^{-30}\,{\rm cm^3/s}$ at 1 MeV up to
$\sim 5\times 10^{-29}\,{\rm cm^3/s}$ at 100 MeV, {\em i.e.} one order of magnitude better than ours.
However, for $p$-wave annihilation, CMB limits degrade up to $\sim 10^{-24}\,{\rm cm^3/s}$ in the
same
mass range (derived assuming a velocity dispersion $\sigma_v =$ 100 km/s). We can roughly convert
our $s$-wave limits in terms of $p$-wave by assuming an isothermal velocity distribution for DM such
that $\sigma_v^{\rm MW}=v_c/\sqrt{2}$, where $v_c\simeq 240$ km/s is the local rotation
velocity \cite{Reid2014}. Therefore, our $s$-wave bounds $\langle \sigma v\rangle_{\rm max}$
rescale to $\langle\sigma v\rangle_{\rm max}(\sigma_v/\sigma_v^{\rm MW})^2$ in terms of $p$-wave,
giving $\sim 3\times 10^{-29}{\rm cm^3/s}$ for $\sigma_v=100$ km/s, {\em i.e.} an improvement by
$\sim5$ orders of magnitude. Finally, our bounds are slightly more stringent than those derived in
gamma-ray studies \cite{Essig2013a}, and less sensitive to the DM halo shape.

To conclude, we have considered for the first time the Voyager $e^{\pm}$ data to derive
constraints on annihilating MeV DM particles (decaying DM in \citesec{app:dec}). Since
Voyager has crossed the heliopause, solar modulation, which prevents MeV CRs to reach space
experiments orbiting the Earth, can be evaded. We used state-of-the-art semianalytic methods to
describe CR propagation, including all relevant
processes. We considered constrained sets of propagation parameters featuring strong (model $A$)
and weak reacceleration (model $B$) to point out an interesting phenomenon: reacceleration may
push $e^\pm$ up to energies higher than $m_\chi$ in the sub-GeV mass range. Thus, GeV data become
constraining also for DM particles in the sub-GeV mass range. We therefore combined the Voyager and
AMS-02 datasets to derive constraints on DM annihilation, getting limits down to
$\langle \sigma v\rangle \sim 10^{-28}{\rm cm^3/s}$ at 10 MeV, quite competitive with respect to
complementary gamma-ray studies, and less dependent on the halo shape. Other complementary CMB
constraints are found more stringent for $s$-wave annihilation but less stringent by about five
orders of magnitude for the $p$-wave. Finally, note that a similar analysis could apply to heavier
DM particles with excited states separated by MeV gaps \cite{Finkbeiner2007}.


\acknowledgments{
  We wish to thank Alan C. Cummings for valuable exchanges about the Voyager 1 data, and Martin
  Winkler for details about the B/C analysis made in Ref.~\cite{Kappl2015a}. We also
  thank the referees for their valuable comments which helped improve the presentation of our
  results. MB acknowledges
  support from the European Research Council (ERC) under the EU Seventh Framework
  Program (FP7/2007-2013)/ERC Starting Grant (agreement n. 278234 | NewDark project led
  by M. Cirelli). JL is partly supported by the OCEVU Labex (ANR-11-LABX-0060), the CNRS program
  {\em D\'efi InPhyNiTi}, and European Union's Horizon 2020 research and innovation program under
  the Marie Sk\l{}odowska-Curie grant agreements No 690575 and
  No 674896 -- in addition to recurrent funding by CNRS and Montpellier University.
  PS is partly supported by the Institut Universitaire de France (IUF).
}

\appendix
\section{Supplemental Material}
\subsection{Hierarchy in the propagation processes}
\label{app:prop}
In \citefig{fig:timescales}, we illustrate the time scales associated with the different
processes at stake during the propagation of $e^\pm$ cosmic rays. All time scales are divided
by the residence time scale in the disk, given by $\tau_{\rm disk}(E)=h\,L/K(E)$, where $h$ is
the half-height of the thin gaseous disk (set to 100 pc) and $L$ is the half-height of the
Galactic magnetic halo. The strong-reacceleration regime
(propagation model $A$) is shown in the left panel, while the weak-reacceleration regime
(model $B$) is displayed in the right panel. In the latter, it is clear that not only does the
ionization loss rate set the full energy loss rate ($1/\tau_l$) in the sub-GeV energy range, but
it also dominates over the other processes. This makes the flux predictions fully fixed by
energy losses and spatial diffusion in this configuration. In the particular case of DM
annihilation in the $e^+e^-$ channel, the prediction of the peak in $E^2\phi_{e^\pm}$ occurring at
the WIMP mass (see \citefig{fig:prims}), which determines the Voyager bounds on
$\langle \sigma v\rangle$, further allows to get rid of spatial diffusion and relies only on
rather well controled local parameters (local dark matter and interstellar gas densities). This
asymptotic peak prediction is therefore insensitive to uncertainties in the propagation parameters
in this weak-reacceleration configuration, which seems to be favored by the most recent B/C data
\cite{Kappl2015a}.

\begin{figure*}[!t]
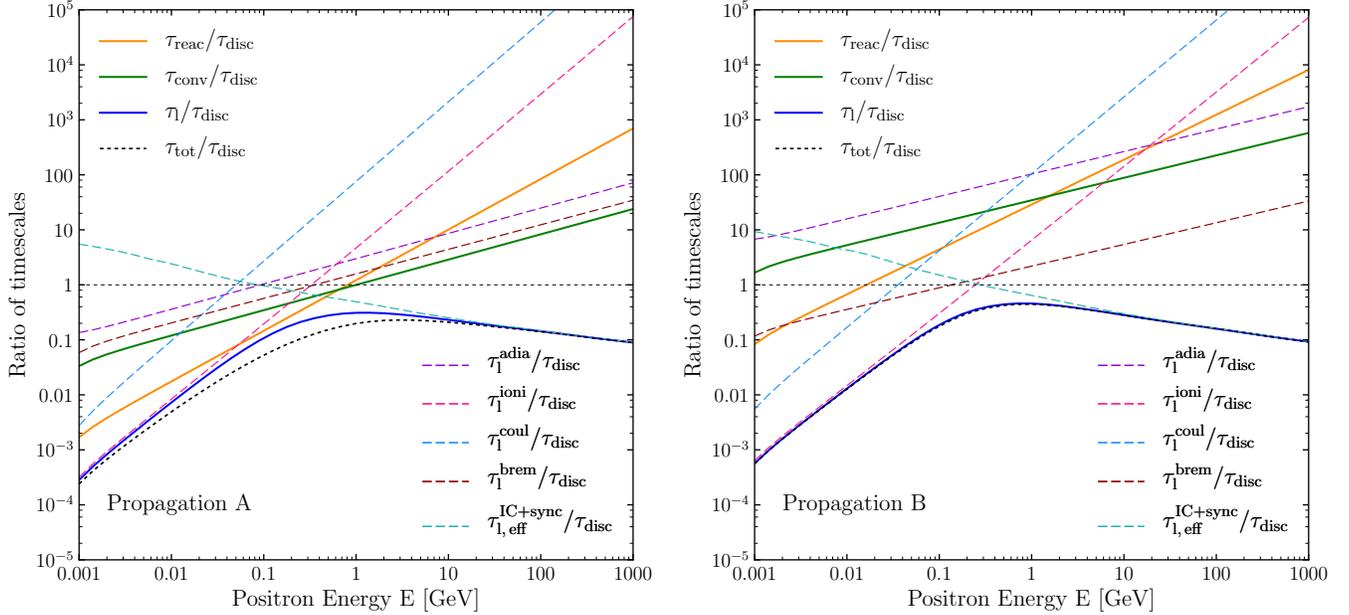

\centering
\includegraphics[width = 0.495\textwidth]{{{ratio_typical_times_A}}}
\includegraphics[width = 0.495\textwidth]{{{ratio_typical_times_B}}}
\caption{\small Left panel: time scales associated with the different propagation processes
  of model $A$. Right panel: the same for model $B$.}
\label{fig:timescales}
\end{figure*}

\subsection{Constraints on dark matter decay}
\label{app:dec}
In this section, we conduct the same analysis as above but for decaying DM particles. The
positron injection rate at position $\vec{x}$ is $\propto \Gamma_\chi\,\rho(\vec{x})/m_\chi$, where
$\Gamma_\chi=1/\tau_\chi$ is the decay rate ($\tau_\chi$ is the DM particle lifetime), and where we
notice the linear dependence in the DM mass density profile, in contrast to the quadratic
dependence that characterizes the annihilation rate. Our lower bounds for the lifetime are reported
in \citefig{fig:decay}, based on the same conservative assumptions as those used to derive
our limits on the annihilation cross section (see right panel of \citefig{fig:limits}).

\begin{figure}[!t]
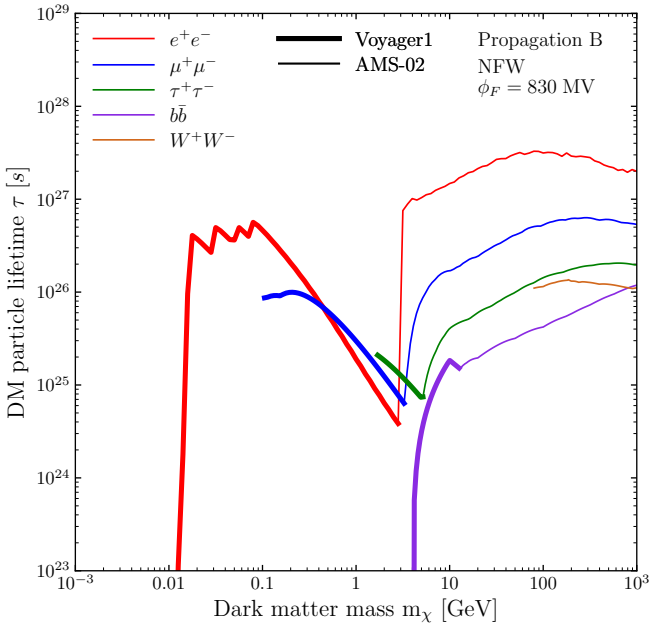

\centering
\includegraphics[width = 0.495\textwidth]{{{taudec_bounds_results_decay}}}
\caption{\small Limits on the DM particle lifetime as a function of the DM particle
    $m_\chi$. This figure assumes the same (conservative) model configuration as in the right
    panel of \citefig{fig:limits}.}
\label{fig:decay}
\end{figure}

\bibliography{biblio_jabref}

\end{document}